\documentclass[twocolumn,aps,prl,showpacs]{revtex4}
\usepackage[T1]{fontenc}
\usepackage[latin9]{inputenc}
\usepackage{amsmath}
\usepackage{amssymb}
\usepackage{stackrel}
\usepackage{graphicx}
\usepackage{wasysym}

\makeatletter
\@ifundefined{textcolor}{}
{%
 \definecolor{BLACK}{gray}{0}
 \definecolor{WHITE}{gray}{1}
 \definecolor{RED}{rgb}{1,0,0}
 \definecolor{GREEN}{rgb}{0,1,0}
 \definecolor{BLUE}{rgb}{0,0,1}
 \definecolor{CYAN}{cmyk}{1,0,0,0}
 \definecolor{MAGENTA}{cmyk}{0,1,0,0}
 \definecolor{YELLOW}{cmyk}{0,0,1,0}
}

\makeatother

\begin{document}

\title{Effects of Spin-Orbit Coupling on Jaynes-Cummings and Tavis-Cummings Models}

\author{Chuanzhou Zhu$^1$, Lin Dong$^1$, and Han Pu$^{1,2}$}

\affiliation{$^{1}$Department of Physics and Astronomy, and Rice Center for Quantum Materials,
Rice University, Houston, TX 77251, USA \\
$^2$Center for Cold Atom Physics, Chinese Academy of Sciences, Wuhan 430071, P. R. China}

\date{\today}
\begin{abstract}
We consider ultracold atoms inside a ring optical
cavity that supports a single plane-wave mode. The cavity field, together with an external coherent laser field, drives a two-photon Raman transition between two internal pseudo-spin states of the atom. This gives rise to an effective coupling
between atom's pseudo-spin and external center-of-mass (COM) motion.
For the case of a single atom inside the cavity, We show how the spin-orbit coupling modifies
the static and dynamic properties of the Jaynes-Cummings (JC) model. In the case of many atoms in thermodynamic limit,
we show that the spin-orbit coupling modifies the Dicke superradiance
phase transition boundary and the non-superradiant normal phase may become reentrant in some regimes.
\end{abstract}


\maketitle

\section{introduction}

The interaction between atomic internal pseudo-spin degrees of freedom
and quantized photon field supported by an optical cavity has long been a focus of the field of cavity quantum electrodynamics (CQED) \cite{cqed}. One of the simplest CQED systems is described by the Jaynes-Cummings (JC) model \cite{sglJC} which concerns the interaction of a single two-level atom and a single-mode cavity field under the rotating wave approximation. Over the past few decades, various techniques have been developed to realize such a system in experiment \cite{sglexp1,sglexp2,sglexp3,sglexp4,sglexp5,sglexp6,sglexp7,sglexp8,sglexp9,sglexp10}, and both the
static \cite{sglexp4,sglexp5,sglexp6} and dynamic \cite{sglexp7,sglexp8}
properties have been explored. The corresponding model with $N$ two-level atoms coupled identically to the single-mode cavity was considered by Dicke \cite{many1} and, later, by Tavis and Cummings \cite{TC}. In the literature, the $N$-atom model with and without the rotating wave approximation are often referred to as the Tavis-Cummings (TC) model and the Dicke model, respectively. It was Dicke who
first suggested to treat all atoms as a single quantum system in
the study of coherent spontaneous radiation process \cite{many1} and proposed what is now called the Dicke states which are a family of correlated $N$-atom states whose spontaneous emission rate scales as $N^2$. In the context of CQED, both the TC and the Dicke
model predict the superradiant phase transition which describes a sudden emergence of macroscopic cavity photon number when the atom-cavity coupling strength exceeds a critical value. Several recent experiments have explored this phoenomenon \cite{manyexp2,manyexp3,manyexp4}. Theoretically,
the Dicke superradiant phase transition in both zero \cite{manyzero5,manyzero6,manyzero7,manyzero8}
and finite \cite{manyfinite} temperatures has been investigated, and
the non-equilibrium physics \cite{manynonequi1,manynonequi2} of the
Dicke model has also been considered.

The advent of cold atoms makes the atomic center-of-mass (COM) motion
no longer negligible, and hence the coupling between the atomic external COM
degrees of freedom and the cavity photon field needs to be considered.
The Bose-Einstein condensate in a CQED system has been realized
in experiments on various platforms \cite{COMexp1,COMexp2}. In this
system, the mutual influence between the atomic COM motion and the
cavity photon field modifies the collective atomic motion \cite{COM3,COM4},
the cavity transmission spectra \cite{COM5}, and can lead to matter wave bistability
\cite{COM6} and multistability behaviors \cite{COM7}, the entanglement
generation \cite{COM8}, etc. In the experimental realization of Dicke model in Ref.~\cite{manyexp2,manyexp3}, the two-level atomic system is formed by two motional states of the atom.

Our purpose in this work is to understand the mutual influence between three degrees
of freedom, including the atomic internal pseudo-spin states, the atomic
COM motion, and the cavity photon field. Due to the fact that the photon field influences both the internal and external states of the atom, an effective spin-orbit coupling (SOC) is realized. Our focus here is to investigate the effects of the SOC on both the JC and the TC models.  
In experimental setups for both Bose \cite{SOCexpB1,SOCexpB2,SOCexpB3,SOCexpB4}
and Fermi \cite{SOCexpF1,SOCexpF2} gases, the SOC
is generated by a pair of counter-propagating coherent laser beams coupling
two hyperfine states of the atom via a two-photon Raman process \cite{Raman}.
Many-body \cite{SOCtheo1,SOCtheo2,SOCtheo3,SOCtheo4,SOCtheo5,SOCtheo6}
and few-body \cite{SOCtheofew1,SOCtheofew2,SOCtheofew3} theories
have been proposed to study the emergence of various quantum phases, and the SOC-induced dynamics \cite{SOCexpF2,PanJW,Shu} has been investigated. In our proposal, we replace one of the Raman laser beams by the cavity field which is {\em dynamically} coupled to the atoms, in the sense that the atomic dynamics provides a back action to the cavity field. Several previous studies have focused on the properties of quantum gases subjected to such dynamic SOC \cite{largymagnetic,lattice,2modes1,2modes2,We1,We2}. In Ref.~\cite{many1}, Dicke already considered the effect of the atomic COM motion on the superradiant emission, although SOC was not explicitly mentioned.

The system we study here, schematically shown in Fig.~\ref{fig0}, is similar to the one we studied in our previous works \cite{We1,We2}, where we have considered the case with one single atom and investigated its energy spectrum, stability properties, and have compared the differences between the semiclassical approach (where the cavity field is treated as coherent field adiabatically following the atomic dynamics) and the full quantum approach. The motivation of the current work is to explore
how the atomic COM motion and the SOC modify the static
and dynamic properties of the JC model and the superradiant phase transition in
the TC model. This paper
is organized as follows. In Sec.~\ref{sec:single-particle-in} we
analytically study the excitation number and the energy dispersion
of a single atom with cavity-assisted SOC in homogeneous
space, and compare these results to the JC model and the classical-laser-induced
SOC. In Sec.~\ref{sec:single-particle-har} we show that the
combination of SOC and a confining trapping potential not only
further modifies the excitation number of the JC model, but also dramatically modifies the spin dynamics.
In Sec.~\ref{sec:n-particle-superradiance} we investigate the Dicke superradiant phase transition of a many-atom system in the thermodynamic
limit, and discuss how the cavity-assisted SOC modifies
the Dicke phase transition boundary. Finally we conclude in Sec. V.

\section{single atom without trap\label{sec:single-particle-in}}

\begin{figure}
\includegraphics[scale=0.66]{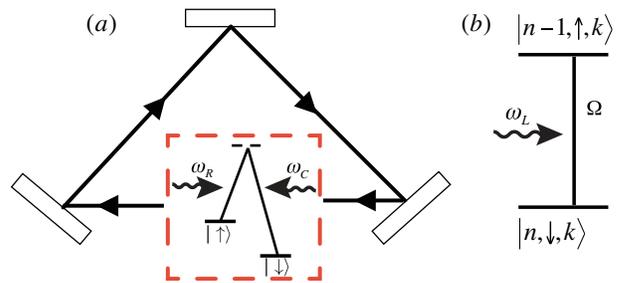}

\caption{(a) Schematic diagram of the spin-orbit coupled system in a ring
cavity. (b) Effective two level model for the scheme in (a), where
a photon field with frequency $\omega_{L}$ induces a transition from
$\left|n,\downarrow,k\right\rangle $ to $\left|n-1,\uparrow,k\right\rangle $,
during which the actual atomic COM momentum $\hat{k}_{\rm{lab}}=\hat{k}+\hat{\sigma}_{z}q_{r}$
changes from $k-q_{r}$ to $k+q_{r}$. }
\label{fig0}
\end{figure}

We consider a single atom with two internal pseudo-spin states, denoted as $\left|\uparrow\right\rangle $
and $\left|\downarrow\right\rangle$, inside a ring cavity, as shown
in Fig.~\ref{fig0}(a). The ring cavity supports a single mode travelling
wave with frequency $\omega_{C}$, and an external light source produces
an additional classical laser beam with frequency $\omega_{R}$. These
two counter-propagating light beams induce a two-photon
Raman transition between the $\left|\uparrow\right\rangle $ and $\left|\downarrow\right\rangle $
states, and simultaneously transfer a recoil momentum of $\pm 2q_{r}$ 
to the atom along the cavity axis which we denote as the $z$-axis. In the lab frame, this cavity-assisted SOC model with the rotating wave approximation is governed by the following Hamiltonian,
\begin{equation}
h_{{\rm lab}}=\frac{\hat{k}^{2}_{\rm lab}}{2m}+\frac{\omega_{0}}{2}\hat{\sigma}_{z}+\frac{{\Omega}e^{2iq_{r}z}}{2}\hat{\sigma}^{+}c+ \frac{{\Omega}e^{-2iq_{r}z}}{2}\hat{\sigma}^{-}c^{\dagger}+\omega_{L}c^{\dagger}c, \label{eq:h_lab_no_trap}
\end{equation}
where $m$ is the atomic mass, $\hat{k}_{\rm lab}$ denotes the atomic COM momentum operator along the cavity axis, $\omega_{0}$
represents the energy difference between $\left|\uparrow\right\rangle $
and $\left|\downarrow\right\rangle $, $c$ and $c^{\dagger}$ are
the cavity photon annihilation and creation operator, respectively, $\omega_{L}=\omega_{C}-\omega_{R}$
describes the frequency difference between two light beams, and $\Omega$
is the single-photon Raman coupling strength. 
Note that Hamiltonian (\ref{eq:h_lab_no_trap}) is written in a frame rotating with the classical laser frequency $\omega_R$.
We will always assume $\omega_L > 0$ in this work, as, otherwise, Hamiltonian (\ref{eq:h_lab_no_trap}) supports no ground state. We have set $\hbar$ equal
to unity for convenience and will choose $\omega_{0}$ as the energy
unit. Here $\hat{\sigma}_{z}$, $\hat{\sigma}^{+}$, and $\hat{\sigma}^{-}$ are defined
as 
\begin{align}
\hat{\sigma}_{z} & =\left|\uparrow\right\rangle \left\langle \uparrow\right|-\left|\downarrow\right\rangle \left\langle \downarrow\right|;\\
\hat{\sigma}^{+} & =\left|\uparrow\right\rangle \left\langle \downarrow\right|; \quad
\hat{\sigma}^{-} =\left|\downarrow\right\rangle \left\langle \uparrow\right|.\nonumber 
\end{align}
Note that for simplicity, we have ignored the atomic COM motion along the two transverse directions perpendicular to the cavity axis, as they are not coupled to the cavity field. It is often more convenient to work in a quasi-momentum frame where the Hamiltonian reads
\begin{equation}
h=\frac{\hat{k}^{2}}{2m}+\frac{q_{r}\hat{k}}{m}\hat{\sigma}_{z}+\frac{\omega_{0}}{2}\hat{\sigma}_{z}+
\frac{\Omega}{2}\left(\hat{\sigma}^{+}c+\hat{\sigma}^{-}c^{\dagger}\right)+\omega_{L}c^{\dagger}c\,.\label{eq:H}
\end{equation}
Here the quasi-momentum frame and the lab frame are connected by a gauge transformation 
$h={\rm U}h_{{\rm lab}}{\rm U}^{\dagger}$
with
\begin{equation}
{\rm U}\equiv{\rm e}^{-iq_{r}z}\left|\uparrow\right\rangle \left\langle \uparrow\right|+{\rm e}^{iq_{r}z}\left|\downarrow\right\rangle \left\langle \downarrow\right|.\label{eq:U}
\end{equation}
Note that in the quasi-momentum frame, $\hat{k}=-i{\partial}/{\partial z}$ represents the COM quasi-momentum operator, which is related to actual atomic momentum $\hat{k}_{\rm lab}$ as $\hat{k}_{\rm{lab}}=\hat{k}+\hat{\sigma}_{z}q_{r}$.
In the following, our discussion will be in the quasi-momentum frame if not otherwise specified.

 
In this homogeneous system, both the quasi-momentum $\hat{k}$ and the excitation
number 
\begin{equation}
\hat{n}=c^{+}c+\left|\uparrow\right\rangle \left\langle \uparrow\right|\,,\label{eq:excitation}
\end{equation}
are conserved. Our model, as described by Hamiltonian (\ref{eq:H}),
can also be effectively viewed as a two level atom coupled by a
photon field with frequency $\omega_{L}$ as shown in Fig.~\ref{fig0}(b).
A coupling is present between the states $\left|n,\downarrow,k\right\rangle $
and $\left|n-1,\uparrow,k\right\rangle $, where $|n_{p}, \sigma ,k \rangle$ denotes a state with $n_{p}$ cavity photons, and the atom in spin-$\sigma$ with quasi-momentum $k$. This spin flipping transition conserves $k$, but the actual atomic COM momentum $\hat{k}_{\rm{lab}}$ changes from $k-q_{r}$ to $k+q_{r}$ as its spin changes from $| \downarrow \rangle$ to $|\uparrow \rangle$ by absorbing a cavity photon. Note that if the photon recoil momentum vanishes, i.e.,
$q_{r}=0$, (which occurs when the cavity field and the external laser beam are co-propagating), the SOC term (the second term on the r.h.s. of Hamiltonian (\ref{eq:H})) is absent, thus the atomic COM motion is completely decoupled from the cavity field. Under this situation, our system is reduced to the conventional JC model after the irrelevant kinetic energy term $\hat{k}^2/(2m)$ in Hamiltonian (\ref{eq:H}) is ignored. In this section,
we investigate the ground state excitation number, and clarify how
the SOC modifies the JC model. We also discuss how the quantization
of the cavity photon field modifies the SOC induced
by two classical lasers.

We choose $\left|n_{p},\sigma,k\right\rangle $ as the basis states. As the excitation number $n$ and atomic quasi-momentum $k$ are good quantum numbers, we can consider the two-dimensional subspace characterized by $n \ge 1$ and atomic momentum $k$ which is spanned by two basis states $\left|n-1,\uparrow,k\right\rangle $
and $\left|n,\downarrow,k\right\rangle $. The Hamiltonian for this subspace is
given by (the subspace for $n=0$ contains only one state $| 0,\downarrow,k \rangle $):
\begin{equation}
h_{n}\left(k\right)=\left[\begin{array}{cc}
h_{n}^{\uparrow}(k) & \frac{\sqrt{n}\Omega}{2}\\
\frac{\sqrt{n}\Omega}{2} & h_{n}^{\downarrow}(k)
\end{array}\right]\,,\label{eq:hn}
\end{equation}
with
\begin{align}
h_{n}^{\uparrow}(k) & =\frac{k^{2}}{2m}+\frac{q_{r}k}{m}+\frac{\omega_{0}}{2}+\left(n-1\right)\omega_{L}\,;\\
h_{n}^{\downarrow}(k) & =\frac{k^{2}}{2m}-\frac{q_{r}k}{m}-\frac{\omega_{0}}{2}+n\omega_{L}\,.\nonumber 
\end{align}
Diagonalizing $h_{n}\left(k\right)$, we obtain two energy dispersions
in this subspace
\begin{align}
\label{eq:Enk}
E_{n\geqslant1}^{\pm}\left(k\right)  =\frac{k^{2}}{2m}\pm\frac{1}{2}\sqrt{\left(\delta^{\rm eff}_{k}\right)^{2}+n\Omega^{2}}+\left(n-\frac{1}{2}\right)\omega_{L} \,,
\end{align}
where 
\begin{equation}
\delta^{\rm eff}_{k} = \delta + 2q_r k/m \,, \label{eq:d_eff_k}
\end{equation}
with $\delta=\omega_{0}-\omega_{L}$ being the bare two-photon detuning for the Raman transition, and the effect of the SOC can be regarded as producing a momentum-dependent effective two-photon detuning $\delta^{\rm eff}_{k}$. To complete the spectrum, we should also include the single dispersion curve in the $n=0$ sector which is given by
\begin{equation}
E_{n=0}^{-}\left(k\right)  =\frac{k^{2}}{2m}-\frac{q_{r}k}{m}-\frac{\omega_{0}}{2} \,. \label{en0}
\end{equation}

\subsection{Ground-State Excitation Number\label{subsec:Ground-State-Excitation-Number}}

In the following, we will first consider the ground state excitation number in this subsection, and then discuss the energy dispersion curve in the next subsection. 

\paragraph{The case with $\delta=0$ ---}
By taking the derivative of $E_{n}^{-}\left(k\right)$, we analytically
obtain one minimum for $E_{0}^{-}\left(k\right)$ at $k=q_{r}$, two
minima for $E_{1\leqslant n<n_{c}}^{-}\left(k\right)$ at $k=\pm q_{r}\sqrt{1-n\Omega^{2}/\left(16E_{r}^{2}\right)}$, and one minimum for $E_{n\geqslant n_{c}}^{-}\left(k\right)$ at
$k=0$, where $n_{c}=\left(4E_{r}/\Omega\right)^{2}$,
and $E_{r}=q_{r}^{2}/\left(2m\right)$ is the photon recoil energy which also characterizes the strength of the SOC. Hence, the energy
minimum $E\left(n\right)$ in each $n$ subspace can be written into
two pieces
\begin{eqnarray}
E\left(n\right) & =n\left(\omega_{0}-\frac{\Omega^{2}}{16E_{r}}\right)-\frac{\omega_{0}}{2}-E_{r}, & n<n_{c}\label{eq:En}\\
E\left(n\right) & =\omega_{0}\left(\sqrt{n}-\frac{\Omega}{4\omega_{0}}\right)^{2}-\frac{\Omega^{2}}{16\omega_{0}}-\frac{\omega_{0}}{2}, & n\geqslant n_{c}\nonumber 
\end{eqnarray}
Finally
we obtain the ground state energy and the corresponding excitation
number by identifying the smallest $E\left(n\right)$ among all $n$'s. 

\begin{figure}
\includegraphics[scale=0.45]{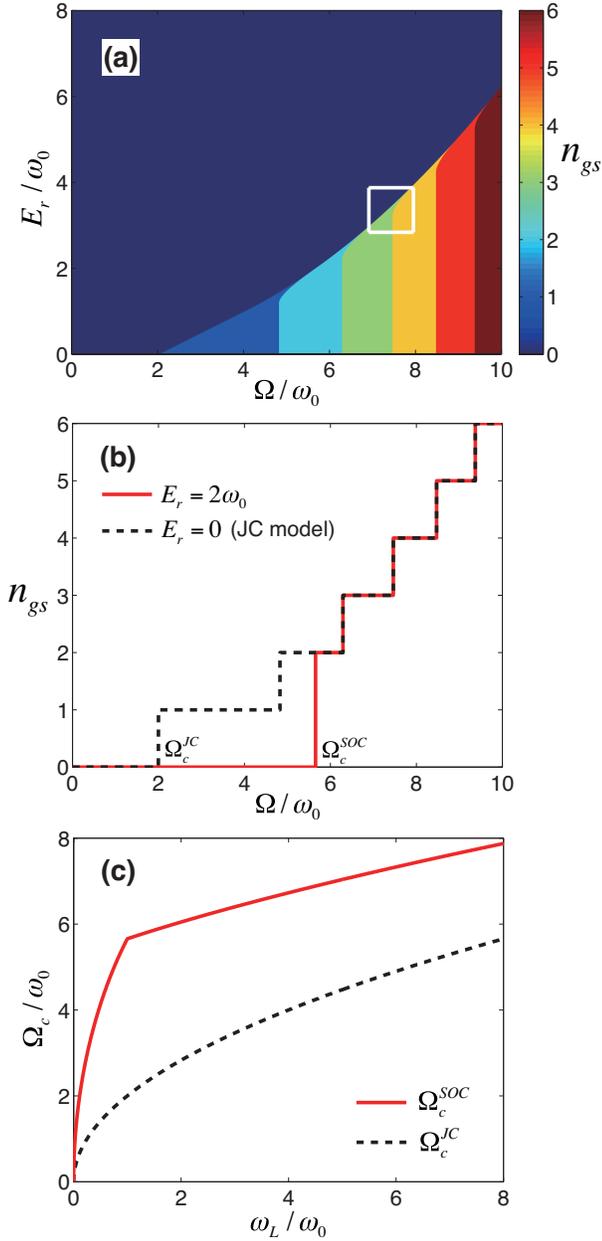}

\caption{(color online) (a) Ground-state excitation number $n_{{\rm gs}}$
of a single atom in a homogeneous space, as a function of Raman coupling
strength $\Omega$ and recoil energy $E_{r}$, with $\omega_{L}=\omega_{0}$.
(b) Ground-state excitation numbers for $E_{r}=2\omega_{0}$ and $E_{r}=0$
(Jaynes-Cummings model) as step functions of $\Omega$ with $\omega_{L}=\omega_{0}$.
(c) The critical Raman coupling strength at which $n{\rm _{gs}}$ jumps from 0 to finite value as a function of $\omega_L/\omega_0$. The red solid line corresponds to $\Omega_{c}^{\text{SOC}}$ with $E_{r}=2\omega_{0}$, and the black dashed line to
$\Omega_{c}^{\text{JC}}$ obtained at $E_{r}=0$.}
\label{fig1}
\end{figure}

Figure~\ref{fig1}(a) presents the ground-state excitation number $n_{{\rm gs}}$
as a function of the Raman coupling strength $\Omega$ and the recoil energy or SOC strength
$E_{r}$. For $E_r=0$, we recover the result for the JC model, where as $\Omega$ increases from zero, $n_{{\rm gs}}$ starts from 0 and increases with steps of one at succeeding critical values of $\Omega$. This is plotted as the black dashed line in Fig.~\ref{fig1}(b). The critical values $\Omega_n$ at which $n_{{\rm gs}}$ jumps from $n$ to $n+1$ can be straightforwardly obtained as  
\begin{align}
\Omega_{n=0} & \equiv\Omega_{c}^{\text{JC}}=2{\omega_{0}}\,;
\label{eq:Omg_c_JC} \\
\Omega_{n\geqslant1} & =2\omega_{0}\left[\left(2n+1\right)+2\left(n^{2}+n \right)^{\frac{1}{2}}\right]^{\frac{1}{2}}\,,\nonumber  
\end{align}
where we have denoted the first critical value as $\Omega_{c}^{\text{JC}}$.

In the presence of SOC (i.e., $E_r \neq 0$), $n_{{\rm gs}}$ still increases in steps at critical values of $\Omega$. However, in comparison to the JC model, there are some key differences. First the parameter regime for $n_{{\rm gs}}=0$ is enlarged, i.e., the first jump where $n_{{\rm gs}}$ changes from 0 to finite occurs at a critical Raman coupling strength $\Omega = \Omega_{c}^{\text{SOC}}>\Omega_{c}^{\text{JC}}$. This is due to the fact that, as can be seen from Eqs.~(\ref{eq:En}) and (\ref{en0}), finite $E_r$ (or $q_r$) reduces the value of $E_{n=0}^-$ more than that of $E_{n \geqslant 1}^-$, which helps to enlarge the $n=0$ regime. 
Here the value of $ \Omega_{c}^{\text{SOC}}$ can be obtained from Eq.~(\ref{eq:En}) as  
\begin{equation}
\Omega_{c}^{\text{SOC}} = \left\{ \begin{array}{ll}   2\left(\omega_{0}+E_{r}\right) \,, & {\rm for}\;\;E_{r}\leqslant\omega_{0} \\ 4\sqrt{\omega_{0}E_{r}} \,, & {\rm for} \;\; E_r > \omega_0 \end{array} \right. \,. \label{eq:Omg_c_SOC}
\end{equation}
Second, at $\Omega = \Omega_{c}^{\text{SOC}}$, $n_{{\rm gs}}$ jumps from 0 to a finite value that is not necessarily equal to 1. An example is shown in Fig.~\ref{fig1}(b) as the red solid line. Third, as $\Omega$ keeps increasing from $\Omega_{c}^{\text{SOC}}$, $n_{{\rm gs}}$ will jump with steps of 1 at exactly the same critical values as in the JC model, because for $\Omega >\Omega_{c}^{\text{SOC}}$, the ground state always occurs at $k=0$ and $h_{n}\left(k=0\right)$ in Eq.~(\ref{eq:En}) takes exactly the same form as that in the JC model.

\paragraph{The case with $\delta \neq 0$ ---} We can proceed in a similar way to obtain results with $\delta \neq 0$. Critical values of $\Omega$ at which $n_{\rm gs}$ jumps can still be found analytically, but the results are too cumbersome to write down explicitly. The main features are not qualitatively different from the previous case with $\delta=0$. In particular, the parameter regime with $n_{\rm gs}=0$ is always enlarged in comparison to the JC model. In other words, we always have $\Omega_{c}^{\text{SOC}}>\Omega_{c}^{\text{JC}}$ at any value of $\delta$, as can be seen in Fig.~\ref{fig1}(c).

\begin{figure}
\includegraphics[scale=0.4]{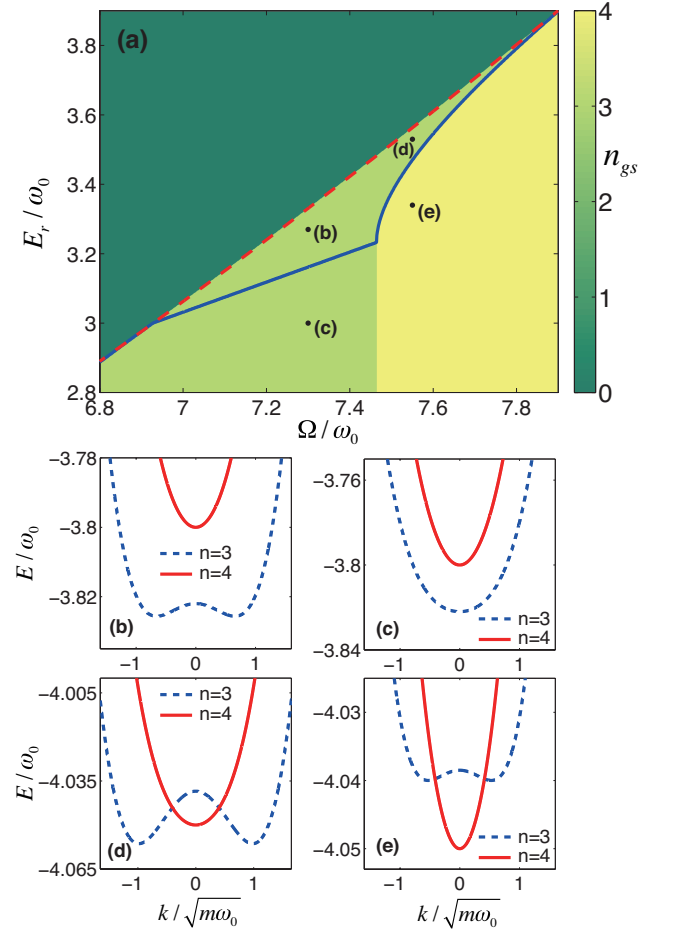}

\caption{(color online) (a) Degenerate-to-nondegenerate transition boundary
and ground state excitation number $n_{{\rm gs}}$ in the parameter space corresponding to the white box in Fig.~\ref{fig1}(a). The
region bounded by the red dash and the blue solid lines features two-fold
degenerate ground state. Outside this region, the ground state is non-degenerate. (b)-(e)
Energy dispersion relations for excitation number $n=3,4$ with $\Omega=7.3\omega_{0}$
and $E_{r}=3.27\omega_{0}$ in (b), $\Omega=7.3\omega_{0}$ and $E_{r}=3\omega_{0}$
in (c), $\Omega=7.55\omega_{0}$ and $E_{r}=3.53\omega_{0}$ in (d),
and $\Omega=7.55\omega_{0}$ and $E_{r}=3.34\omega_{0}$ in (e). These
parameters correspond to points (b)-(e) in (a).}
\label{fig2}
\end{figure}

\subsection{Energy Dispersion and Degeneracy }

We now discuss the energy dispersion curve and ground state degeneracy, which are determined by Eqs.~(\ref{eq:Enk}) and (\ref{en0}). For the Raman spin-orbit coupling induced by two classical laser beams
whose Hamiltonian is given by 
\begin{equation}
h_{{\rm cl}}=\frac{\hat{k}^{2}}{2m}+\frac{q_{r}\hat{k}}{m}\hat{\sigma}_{z}+\frac{\delta}{2}\hat{\sigma}_{z}+\frac{\Omega_{{\rm cl}}}{2}\hat{\sigma}^{+}+\frac{\Omega_{{\rm cl}}}{2}\hat{\sigma}^{-} \,,\label{eq:h_cl}
\end{equation}
it is well known \cite{SOCtheo6} that,
for $\delta=0$, the energy dispersion exhibits a single minimum
when $E_{r}\leqslant\Omega_{\rm{cl}}/4$, and two double minima
when $E_{r}>\Omega_{\rm{cl}}/4$.

In our system with quantized light field, things become more complicated. An example is shown in Fig.~\ref{fig2}.
In Fig.~\ref{fig2}(a), whose parameter space corresponds to that represented by the white box in Fig.~\ref{fig1}(a), the background color represents the value of ground state excitation number $n_{\rm gs}$. The region bounded by the red dashed and the blue solid lines has two-fold degenerate ground state, while the other region features nondegenerate ground state. The two lowest dispersion curves of 4 points labelled by (b-e) in Fig.~\ref{fig2}(a) are plotted in Fig.~\ref{fig2}(b-e). 

As $E_{r}$ decreases from point (b) to (c), the $n=3$ dispersion
transforms from two double minima to a single
minimum, and the ground state changes from
degenerate to non-degenerate. This process is similar to what happens
in the classical-laser-induced SOC, since the ground
state always stays in $n=3$ dispersion. 

As $E_{r}$ decreases from point (d) to (e), the $n=3$ dispersion curve always possess two minima, but the ground state changes from degenerate to nondegenerate as the ground state excitation number $n_{\rm gs}$ jumps from 3 to 4. Hence, this process is a unique feature of the cavity-assisted SOC.

%
%

\section{single atom in harmonic trap\label{sec:single-particle-har}}

In the absence of the trapping potential, the atomic quasi-momentum is conserved. For a fixed quasi-momentum $k$, Hamiltonian (\ref{eq:H}) is the same as that for the JC model, and the SOC term is to effectively shift the atomic transition frequency $\omega_0$ by a momentum-dependent amount of $2q_r k/m$, or equivalently to give rise to a momentum-dependent detuning $\delta^{\rm eff}_{k}$ defined in Eq.~(\ref{eq:d_eff_k}). When a trapping potential is present, $k$ will no longer be a good quantum number, and different quasi-momentum components will therefore be coupled together. This is the situation we are now going to investigate. Specifically, we will consider the presence of a harmonic trap with trapping frequency $\omega_t$. The total Hamiltonian is now 
\begin{equation}
h_{t}=h+\frac{1}{2}m\omega_{t}^{2}z^{2},\label{eq:ht}
\end{equation}
where $h$ is given in Eq.~(\ref{eq:H}).

\subsection{Ground-State Excitation Number}

We first consider the ground state excitation number. Note that, even in the presence of the trapping potential, the excitation number $n$, defined in Eq.~(\ref{eq:excitation}), remains a good quantum number.
Note that Hamiltonian (\ref{eq:ht}) can be represented by ladder operators $a$ and $a^{\dagger}$ of the harmonic oscillator as
\begin{equation}
\begin{split}
h_{t}=&\omega_{t}a^{\dagger}a
+iq_{r}\sqrt{\frac{\omega_{t}}{2m}}
\left(a^{\dagger}-a\right)\sigma_{z} \\
&+\frac{\omega_{0}}{2}\hat{\sigma}_{z}+
\frac{\Omega}{2}\left(\hat{\sigma}^{+}c+\hat{\sigma}^{-}c^{\dagger}\right)+\omega_{L}c^{\dagger}c
,\label{eq:ht_photon}
\end{split}
\end{equation}
where we have used $\hat{k}=i\sqrt{m\omega_{t}/2}\left(a^{\dagger}-a\right)$ and made an energy shift of $\omega_{t}/2$.
We obtain the ground state through the exact
diagonalization approach by expanding the Hamiltonian (\ref{eq:ht_photon}) onto the basis states $\left|n_{p},\sigma,q\right\rangle$,
where $\left|n_{p}\right\rangle$ is the photon Fock state, $\left|\sigma\right\rangle$ is the atomic spin state, and $\left|q\right\rangle$ represents
the phonon Fock state of the harmonic oscillator defined by $a^{\dagger}a\left|q\right\rangle=q\left|q\right\rangle$.
A sufficiently large cutoffs for $n_{p}$ and $q$ are chosen in the calculation.

\begin{figure}
\includegraphics[scale=0.3]{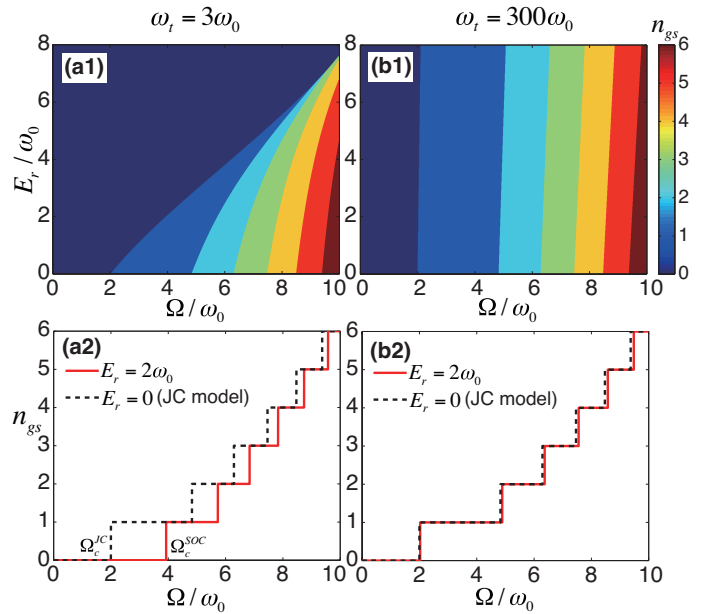}

\caption{(color online) (a1)(b1) Ground-state excitation number $n_{{\rm gs}}$
of a single atom in harmonic trap with trap frequency $\omega_{t}=3\omega_{0},300\omega_{0}$,
as a function of Raman coupling strength $\Omega$ and recoil energy
$E_{r}$. (a2)(b2) The $n_{{\rm gs}}$ for $E_{r}=2\omega_{0}$ and
$E_{r}=0$ (Jaynes-Cummings model) as step functions of $\Omega$
with $\omega_{t}=3\omega_{0},300\omega_{0}$. Here we consider $\omega_{L}=\omega_{0}$.}
\label{fig3}
\end{figure}

Figure~\ref{fig3}(a1) shows the ground-state excitation number $n_{{\rm gs}}$
as a function of Raman coupling strength $\Omega$ and recoil energy
$E_{r}$ for $\delta=0$ in the presence of a relatively weak harmonic trap with trap frequency $\text{\ensuremath{\omega}}_{t}=3\omega_{0}$.
Compared to the previous result without the trap as shown in Fig.~\ref{fig1}(a), here
the boundaries between different $n_{{\rm gs}}$ are bent curves instead
of straight lines. 
Figure \ref{fig3}(a2) shows $n_{{\rm gs}}$ as a function of $\Omega$ for two values of $E_r$. The case with $E_r=0$ corresponds to the absence of SOC and our model is reduced to the JC model. At finite $E_r$, the SOC term shifts the values of the critical Raman coupling strength at which $n_{\rm gs}$ jumps. 
In addition, $\Omega_{c}^{\text{SOC}}>\Omega_{c}^{\text{JC}}$ is
still satisfied for any $\omega_{L}\geqslant 0$ as in the previous case of homogeneous space. 

In Fig.~\ref{fig3}(b1) and (b2), we plot the $n_{\rm gs}$ for a relatively strong harmonic trap with  $\text{\ensuremath{\omega}}_{t}=300 \omega_0$. In this case, we find that the results for finite $E_r$ are not very different from the JC model results as long as $E_r \ll \omega_t$. This can be intuitively understood as follows. In the presence of a very strong trapping potential, the effect of photon recoil, and hence that of the SOC, becomes less important. This is analogous to the Lamb-Dicke limit in the context of ion trapping, in which the coupling between the ion's internal dynamics and its motional dynamics induced by an external light field is suppressed by a strong confining potential.

\subsection{Spin Dynamics}
In the JC model, when a cavity field with definite photon number (i.e., a cavity Fock state) is coupled to the two-level atom, the ensuing spin dynamics is described by the well-known Rabi oscillation, where the oscillation frequency is determined by the coupling strength $\Omega$ and the detuning $\delta$. In our model, the trapping term couples different quasi-momentum states, and each quasi-momentum state experiences a momentum-dependent effective Raman detuning $\delta^{\rm eff}_{k}$. The resulting spin dynamics becomes much more complicated.

To investigate the spin dynamics in our model, we consider a specific initial state $|\psi(0) \rangle = |n_p, \downarrow, q=0 \rangle$ in the lab frame, where the atom is prepared in the $| \downarrow \rangle$ state and the ground state of the harmonic trap, and the cavity field is in a Fock state with $n_p$ photons. This confines the system dynamics within the subspace characterized by excitation number $n=n_p$. Within this subspace, the Hamiltonian in Eq.~(\ref{eq:ht}) takes the form as (after neglecting a dynamically irrelevant constant term)
\begin{equation}
h_{t}\left(n_{p}\right)=\frac{\hat{k}^{2}}{2m}+\frac{q_{r}\hat{k}}{m}\hat{\sigma}_{z}+\frac{\delta}{2}\hat{\sigma}_{z}+\frac{\Omega_{{\rm cl}}}{2}\hat{\sigma}^{+}+\frac{\Omega_{{\rm cl}}}{2}\hat{\sigma}^{-} +\frac{1}{2} m\omega_t^2 z^2\,,\label{eq:h_t_np}
\end{equation}
where $\Omega_{\rm cl} \equiv \Omega \sqrt{n_p}$,
and $\hat{\sigma}_{z}$, $\hat{\sigma}^{+}$, and $\hat{\sigma}^{-}$ are re-defined as
\begin{align}
\hat{\sigma}_{z} & =\left| {n_{p}-1}, \uparrow\right\rangle \left\langle {n_{p}-1}, \uparrow\right|-
\left| {n_{p}}, \downarrow\right\rangle \left\langle {n_{p}}, \downarrow\right|;\\
\hat{\sigma}^{+} & =\left| {n_{p}-1}, \uparrow\right\rangle \left\langle {n_{p}}, \downarrow\right|; \quad
\hat{\sigma}^{-} =\left| {n_{p}}, \downarrow\right\rangle \left\langle {n_{p}-1}, \uparrow\right|.\nonumber 
\end{align}
Note that this Hamiltonian is mathematically equivalent to the Hamiltonian describing a spin-orbit coupled atom where the SOC is generated by two classical Raman laser beams (see Eq.~(\ref{eq:h_cl})) \cite{SOCexpB1,SOCexpB2,SOCexpB3,SOCtheo6,SOCexpF1,SOCexpF2}. As a consequence, the result presented below is also valid in that context. In the classical laser context, the corresponding spin dynamics has been studied in \cite{SOCexpF2,Shu}, whereas we focus on the effects of the photon recoil on the Rabi oscillation in cases of different trapping strengths.

We solve the time-dependent Schr\"{o}dinger equation numerically to find the state vector $|\psi(t) \rangle$ starting from the initial state $|\psi(0) \rangle$, we then calculate the probability of finding the atom in $| \uparrow \rangle$:
\begin{equation}
P_{\uparrow}\left(t\right)\equiv\stackrel[q=0]{\infty}{\sum}\left|\left\langle n_{p}-1,\uparrow,q\right|\left.\psi\left(t\right)\right\rangle \right|^{2}.\label{eq:P_up}
\end{equation}
\begin{figure}
\includegraphics[scale=0.36]{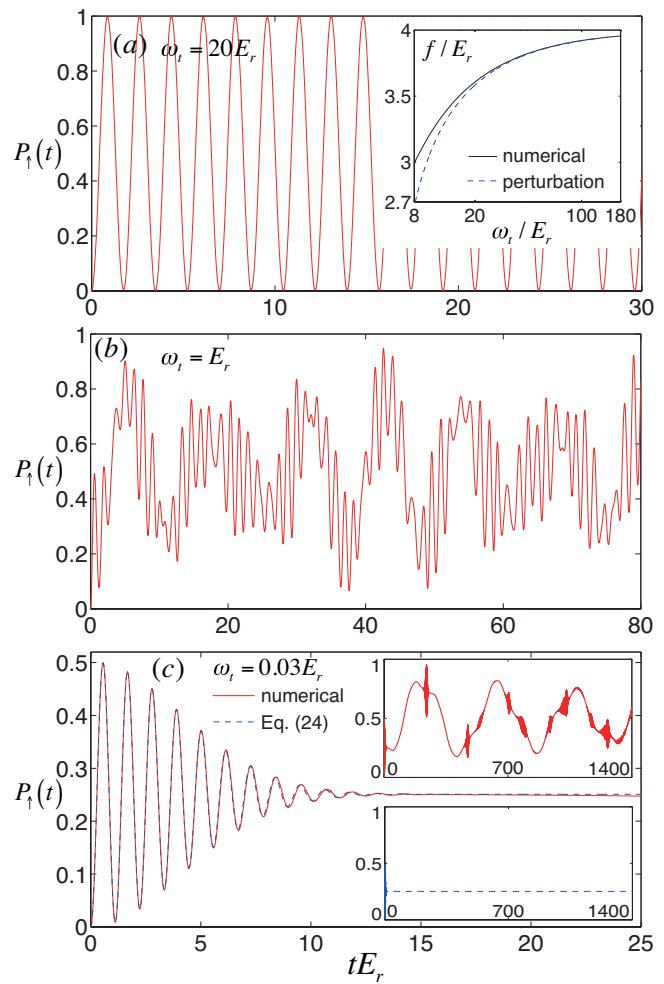}
\caption{(color online) (a)-(c) Time evolution of the spin-up probability $P_{\uparrow}\left(t\right)$
of a single atom in a harmonic trap with $\omega_{t}=20E_{r}$, $E_{r}$, and $0.03E_{r}$.
The system is initially prepared in the harmonic oscillator ground
state with spin down. The inset of (a) plots $f\left(\omega_{t}\right)$,
the oscillation frequency of $P_{\uparrow}\left(t\right)$, as a function
of $\omega_{t}$ in the strong trap regime, where the black solid
line depicts the numerical result obtained by Fourier analysis and
the blue dashed line depicts the analytical result from the perturbation
theory [Eq.~(\ref{eq:f_pertur1})], and a logarithmic scale is used for the horizontal axis.
In (c) for the weak trap regime, the red solid line shows the numerical result, and the blue dashed line shows the analytic result in Eq.~(\ref{eq:P_up_no_trap}) where the coupling between different momentum spaces is neglected. The corresponding long time evolutions of $P_{\uparrow}\left(t\right)$ are shown in the inset of (c).
Other parameters: $\Omega_{{\rm cl}}=4E_{r}$, $\delta=0$.}
\label{figt}
\end{figure}

Examples of spin dynamics are plotted in Fig.~\ref{figt}(a-c), which represent the $\delta=0$ cases for a strong, an intermediate, and a weak trap, respectively, where the trap strength is measured against $E_r$.

\paragraph{Strong Trap ---} As we discussed in the previous subsection, in the presence of a strong trap with $\omega_t \gg E_r$, the system is in the Lamb-Dicke regime where the effect of SOC may be regarded as a small perturbation. The corresponding spin dynamics shown in Fig.~\ref{figt}(a) is accurately described by a sinusoidal oscillation as
\begin{equation}
P_{\uparrow}\left(t\right)=\sin^{2}\left[\frac{
f\left(\omega_{t}\right)
}{2}t\right]\,,\label{eq:P_f_omgz}
\end{equation}
where $f\left(\omega_{t}\right)$ denotes the oscillation frequency which depends on the trap frequency $\omega_{t}$. In the limit of $\omega_t \rightarrow \infty$, the JC model result is recovered as the oscillation frequency 
$f\left(\omega_{t}\rightarrow\infty\right)=\Omega_{\rm cl}$, with the Rabi frequency $\Omega_{\rm cl}=4E_{r}$ in this example. For large but finite $\omega_t$, the oscillation frequency $f\left(\omega_{t}\right)$ deviates away from this value. By treating the SOC term as a small perturbation, we can analytically obtain the oscillation frequency as
\begin{equation}
f\left(\omega_{t}\right)=\Omega_{{\rm cl}}-\frac{2E_{r}\Omega_{{\rm cl}}}{\omega_{t}}-\frac{2E_{r}\Omega_{{\rm cl}}^{3}}{\omega_{t}\left(\omega_{t}^{2}-\Omega_{{\rm cl}}^{2}\right)}\,.\label{eq:f_pertur1}
\end{equation}  
Details of this derivation can be found in Appendix~\ref{sec:perturbation-theory-for}. In the inset of Fig.~\ref{figt}(a), we compare the spin oscillation frequency obtained from the numerical calculation (black solid line) and the analytic result of Eq.~(\ref{eq:f_pertur1}) (blue dashed line), and find excellent agreement for large $\omega_t$. 
 

\paragraph{Weak Trap ---} An example of weak trap with $\omega_t \ll E_r$ is presented in Fig.~\ref{figt}(c), where the short- and long-time behaviours are plotted in the main figure and the insets, respectively. For short-time scale, the system exhibits a damped oscillation. This damped oscillation can be intuitively understood as follows. The initial COM wave function of the atom is a Gaussian (the ground state of the harmonic oscillator), which in the (quasi-)momentum space can be written as 
\begin{equation}
\phi_{0}\left(k\right)=\left(\pi m\omega_{t}\right)^{-\frac{1}{4}}e^{-\frac{(k-q_r)^{2}}{2m\omega_{t}}}\,.
\end{equation}
For such a weak trap, and for short time scale, we can neglect the trap-induced coupling between different momentum components. Then each momentum component exhibits Rabi oscillation, such that for a given quasi-momentum $k$ we have 
\begin{equation}
p_\uparrow ({t,k})=\frac{\Omega_{{\rm cl}}^{2}}{\Omega_{{\rm cl}}^{2}+\left(\delta_{k}^{{\rm eff}}\right)^{2}}\sin^{2}\left(\frac{1}{2}\sqrt{\Omega_{{\rm cl}}^{2}+\left(\delta_{k}^{{\rm eff}}\right)^{2}}t\right)\,,
\end{equation}
where $\delta_{k}^{{\rm eff}}={2q_{r}k}/{m}$ is the effective two-photon detuning for the given momentum component $k$.
Integrating over all the momentum components, we have
\begin{equation}
P_{\uparrow}\left(t\right)=\int dk \, |\phi_0(k)|^2p_\uparrow ({t,k}) \, .\label{eq:P_up_no_trap}
\end{equation}
In the main figure of Fig.~\ref{figt}(c), the red solid line represents the result obtained from the numerical calculation and the blue dashed line the result based on Eq.~(\ref{eq:P_up_no_trap}). Both results agree with each other very well. The damping of the oscillation arises from the dephasing effect, as different momentum components oscillate at different frequencies due to the momentum-dependent effective detuning $\delta_{k}^{\rm eff}$. 

For time scales on the order of or longer than $1/\omega_t$, the assumption underlying Eq.~(\ref{eq:P_up_no_trap}) that different momentum components behave independently is no longer valid. The numerically obtained long-time result and the one based on Eq.~(\ref{eq:P_up_no_trap}) are plotted in the insets of Fig.~\ref{figt}(c). Significant discrepancies can be seen. In particular, Eq.~(\ref{eq:P_up_no_trap}) predicts a featureless flat line: once the dephasing occurs, $P_\uparrow$ no longer oscillates and stays constant. But the full numerical result shows that, due to the momentum components coupling induced by the trapping potential, the long-time behaviour of the system can be quite rich.

\section{superradiance in thermodynamic limit\label{sec:n-particle-superradiance}}
So far, we have been focusing on the properties of a single atom.
In this section, we consider a system where the single mode cavity photon
field is coupled to many atoms in thermodynamic
limit. We neglect the bare interactions between atoms. However, as each atom influences the whole photon field
which back acts on the other atoms, the photon field induces
an effective coupling between atoms. When the atomic COM motion is neglected, our model reduces to the TC model. One of the most well-known
many-body effects in this model is the Dicke
superradiant phase transition \cite{manyzero5,manyzero6,manyzero7,manyzero8,manyfinite}.
Here we investigate how the COM
degree of freedom and the SOC affect the Dicke phase transition.

We consider a canonical ensemble where $N$ atoms inside the cavity are confined within a box with
volume $V$. In the thermodynamic limit,
both $N$ and $V$ are taken to be infinity 
but the number density $\rho=N/V$ is finite. The Hamiltonian of this
system is given by
\begin{equation}
H=\omega_{L}c^{\dagger}c+\sum_{j=1}^{N}\hat{h}_{j}\,,
\end{equation}
with the Hamiltonian for the $j$th atom 
\begin{equation}
\hat{h}_{j}=\frac{\hat{\mathbf{k}}_{j}^{2}}{2m}+\frac{q_{r}\hat{k}_{zj}}{m}\sigma_{z}^{j} +\frac{\omega_{0}}{2}\sigma_{z}^{j} +\frac{\tilde{\Omega}}{2\sqrt{N}} \left(\sigma_{j}^{+}c+\sigma_{j}^{-}c^{\dagger} \right)\,,
\end{equation}
where $\tilde{\Omega}=\sqrt{N}\Omega$ is the rescaled Raman coupling strength,
and $\hat{\mathbf{k}}_{j}$ is the three dimensional quasi-momentum operator for the $j$th atom.

To investigate the thermodynamic phase transition at temperature $T$, we take a similar approach as in Ref.~\cite{manyfinite}
in which the Dicke phase transition in the TC model is investigated.
The canonical partition function $Z={\rm Tr}\left(e^{-\beta H}\right)$
with $\beta=1/\left(k_{B}T\right)$ can be calculated as
\begin{equation}
Z=\frac{V^{N}}{\left(2\pi\right)^{3N}}
\int\frac{d^{2}\alpha}{\pi}
\prod_{j=1}^{N}
\left(\int d\mathbf{k}_{j}
\sum_{\sigma_{j} = \uparrow,\downarrow}\right)
\left\langle \Psi\right|e^{-\beta H}\left|\Psi\right\rangle, \label{eq:Z}
\end{equation}
where we have chosen the states
\begin{equation}
\left|\Psi\right\rangle =\left|\alpha\right\rangle \prod_{j=1}^{N}\left|\mathbf{k}_{j}\right\rangle \left|\sigma_{j}\right\rangle \label{eq:basis}
\end{equation}
as our basis states to evaluate the trace. Here $\left|\alpha\right\rangle $ is the photon coherent state, i.e., the eigenstate of photon annihilation operator such that $ c|\alpha \rangle = \alpha |\alpha \rangle$,
$\left|\mathbf{k}_{j}\right\rangle $ is
the quasi-momentum eigenstate for the $j$th atom, and $\left|\sigma_{j}\right\rangle $ ($\sigma_{j}=\uparrow,\downarrow$) is the
eigenstate of $\sigma_{z}^{j}$ for the $j$th atom. By using the condition $N\rightarrow\infty$,
we obtain
\begin{equation}
\left\langle \alpha\right|e^{-\beta H}\left|\alpha\right\rangle =\exp\left[-\beta\left(\omega_{L}\left|\alpha\right|^{2}+\sum_{j=1}^{N}\hat{h}_{j}^{\alpha}\right)\right]\,,
\end{equation}
where
\begin{equation}
\hat{h}_{j}^{\alpha}=\frac{\hat{\mathbf{k}}_{j}^{2}}{2m}+\frac{q_{r}\hat{k}_{zj}}{m}\sigma_{z}^{j} +\frac{\omega_{0}}{2}\sigma_{z}^{j} +\frac{\tilde{\Omega}}{2\sqrt{N}}\left(\sigma_{j}^{+}\alpha+\sigma_{j}^{-}\alpha^{*}\right)\,.
\label{ha}
\end{equation}
As the summation over spin and integral over momentum in Eq.~(\ref{eq:Z}) are independent for different atoms, the
partition function can be simplified as
\begin{equation}
Z=\int\frac{d^{2}\alpha}{\pi}e^{-\beta\omega_{L}\left|\alpha\right|^{2}}\left[\frac{V}{\left(2\pi\right)^{3}}\int d\mathbf{k}\left(e^{-\beta\epsilon^{+}}+e^{-\beta\epsilon^{-}}\right)\right]^{N},\label{eq:Z2}
\end{equation}
where 
\begin{equation}
\epsilon^{\pm}=\frac{\mathbf{k}^{2}}{2m}\pm\sqrt{\left(\frac{q_{r}k_{z}}{m}
+\frac{\omega_{0}}{2}\right)^{2}+\left(\frac{\tilde{\Omega}}{2}\right)^{2}\frac{\left|\alpha\right|^{2}}{N}}
\end{equation}
are the eigenvalues of $\hat{h}_{j}^{\alpha}$ in Eq.~(\ref{ha}). Integrating over
the complex angle of $\alpha$ and $x,y$ components of $\mathbf{k}$
in (\ref{eq:Z2}), and letting $u=\frac{\left|\alpha\right|^{2}}{N}$,
we can rewrite the partition function as
\begin{equation}
Z=C_{1}\int_{0}^{\infty}du\exp\left\{ N\left[F(u)\right]\right\} \,,\label{eq:Z4}
\end{equation}
with constant $C_{1}=N\left(\frac{mV}{4\pi^{2}\beta}\right)^{N}$ and 
\begin{align}
F(u) &= -\beta\omega_{L}u+\log S\left(u\right)\,,\\
S\left(u\right) & =2\int dk_{z}\exp\left(-\frac{\beta k_{z}^{2}}{2m}\right)\cosh\xi\left(k_{z},u\right)\,,\\
\xi\left(k_{z},u\right) & =\beta\sqrt{\left(\frac{q_{r}k_{z}}{m}
+\frac{\omega_{0}}{2}\right)^{2}+\left(\frac{\tilde{\Omega}}{2}\right)^{2}u}\,.
\end{align}
The Laplace's method \cite{manyfinite} is used to deal with the integral
over $u$ in Eq. (\ref{eq:Z4}). For $N\rightarrow\infty$, this yields
\begin{equation}
Z=C_{2}\max_{u\in\left[0,\infty\right)}\exp\left\{ N\left[F(u)\right]\right\} ,\label{eq:Z5}
\end{equation}
where $C_{2}$ is a constant and we denote that the maximum of $F(u)$ is reached
at $u=u_{0}$. We numerically obtain a $u_{0}\geqslant 0$
by taking the first and second order derivatives of $F(u)$, and it is straightforward to show that $u_{0}$ is actually
the normalized photon number
\begin{equation}
u_{0}=\frac{\left\langle c^{\dagger}c\right\rangle }{N}\,,
\end{equation}
where $\left\langle c^{\dagger}c\right\rangle/N>0$ corresponds to the superradiant phase with a macroscopic
photon excitation appearing in the thermodynamic limit; and $\left\langle c^{\dagger}c\right\rangle/N=0$
corresponds to the normal phase. 

\begin{figure}
\includegraphics[scale=0.42]{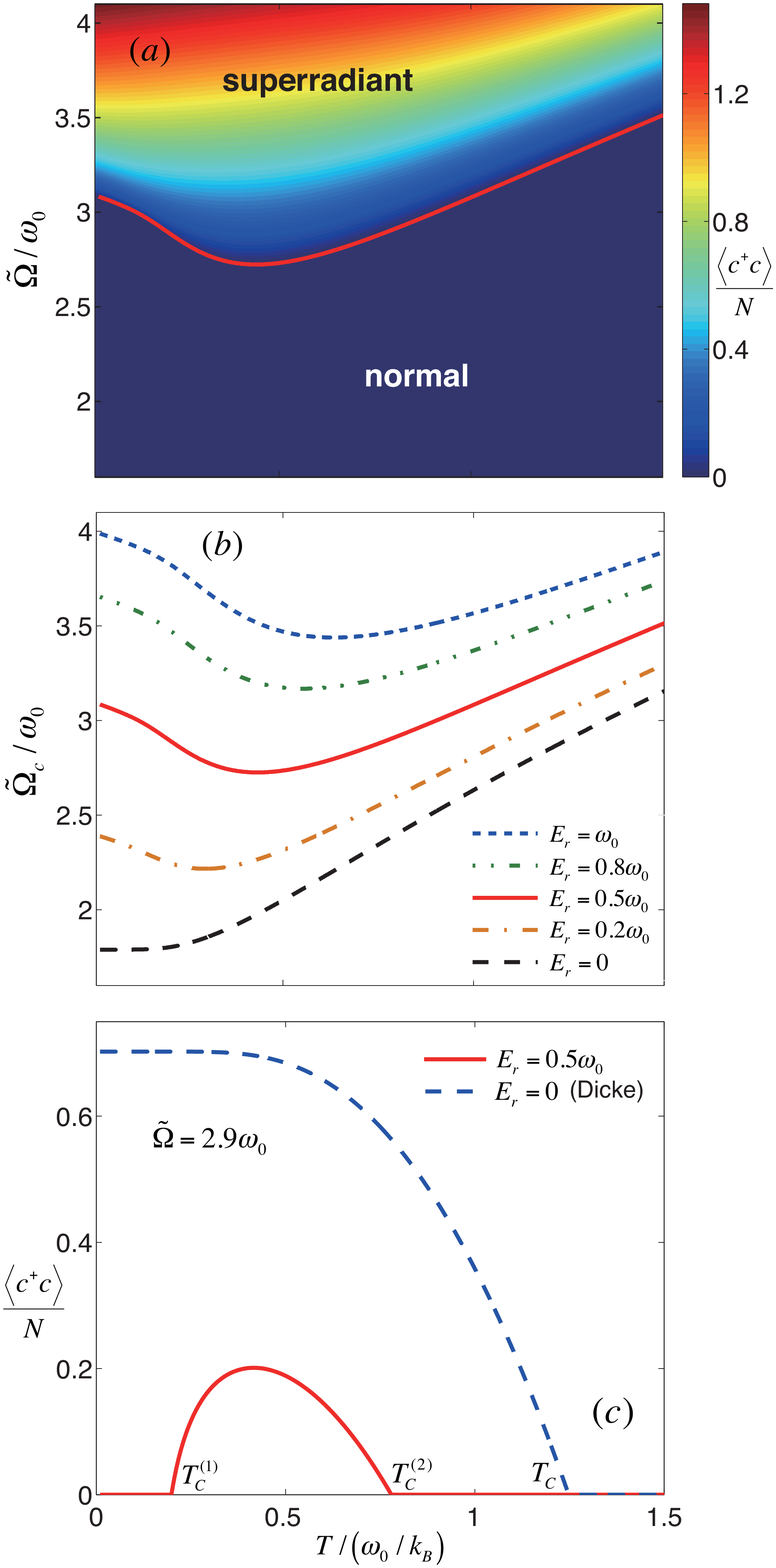}

\caption{(color online) (a) Normalized photon number $\left\langle c^{\dagger}c\right\rangle /N$
as a function of temperature $T$ and effective Raman coupling strength
$\widetilde{\Omega}$ with $E_{r}=0.5\omega_{0}$, where $\left\langle c^{\dagger}c\right\rangle $
is the average photon number and $N$ is the atom number. Here $\left\langle c^{\dagger}c\right\rangle /N>0$
corresponds to the superradiant phase and $\left\langle c^{\dagger}c\right\rangle /N=0$
corresponds to the normal phase. 
(b) Normal-Superradiant Phase boundary
in $T-\widetilde{\Omega}$ plane for $E_{r}/\omega_{0}=0,0.2,0.5,0.8,1$.
(c) $\left\langle c^{\dagger}c\right\rangle /N$
as a function of $T$ for $\widetilde{\Omega}=2.9\omega_{0}$ with
$E_{r}/\omega_{0}=0,0.5$.
We take $\omega_{L}=0.8\omega_{0}$ in these figures.}
\label{fig5} 
\end{figure}

Figure \ref{fig5}(a) shows $\left\langle c^{\dagger}c\right\rangle/N$ as a function of the temperature
$T$ and the rescaled Raman coupling strength $\tilde{\Omega}$ with
the SOC strength $E_{r}=0.5\omega_{0}$. The red solid line in the figure represents the critical coupling strength $\tilde{\Omega}_c$ (i.e., the phase boundary): Above this line, we have $\left\langle c^{\dagger}c\right\rangle/N>0$ and the system is in 
the superradiant phase; and below this line, $\left\langle c^{\dagger}c\right\rangle/N=0$ which
corresponds to the normal phase. In Fig.~\ref{fig5}(b), we plot $\tilde{\Omega}_c$ as a function of $T$ for several different values of $E_r$. As in the previous single-atom case, we recover the usual TC model when $E_r=0$ (bottom curve in Fig.~\ref{fig5}(b)). For the TC model, $\tilde{\Omega}_c$ is a monotonically increasing function of $T$, and $\tilde{\Omega}_c =2\sqrt{ \omega_0 \omega_L}$ at $T=0$. For finite $E_r$, $\tilde{\Omega}_c$ is larger than the corresponding value in the TC model. In other words, in the presence of the SOC, the regime of normal phase is enlarged, which is consistent with the single-atom property that the SOC enlarges the $n=0$ regime with no photons, as shown in Eqs.~(\ref{eq:Omg_c_JC}) and (\ref{eq:Omg_c_SOC}), and Fig.~\ref{fig1}. The upward shift of $\tilde{\Omega}_c$ at finite $E_r$ is more pronounced at lower temperature. This may not be surprising as, at lower temperature, the average atomic COM kinetic energy is lower and hence the photon recoil plays a more significant role. This temperature dependent shift leads to another important feature brought by the SOC: $\tilde{\Omega}_c$ is no longer a monotonic function of $T$, as can be easily seen in Fig.~\ref{fig5}(b), and reaches the minimum value at a finite temperature. 

A consequence of the nonmonotonic behaviour of $\tilde{\Omega}_c$ is that the normal phase may become reentrant as the temperature varies. This is depicted in Fig.~\ref{fig5}(c), where we plot $\left\langle c^{\dagger}c\right\rangle/N$ as a function of $T$ with $\tilde{\Omega}=2.9\omega_{0}$ for 
$E_{r}=0.5\omega_{0}$ (red solid line) and $E_{r}=0$ (blue dashed line). For the TC model ($E_{r}=0$), the system is in the superradiant phase at sufficiently low temperature when $\tilde{\Omega} > 2\sqrt{ \omega_0 \omega_L}$ (as is the case shown in Fig.~\ref{fig5}(c)) with finite $\left\langle c^{\dagger}c\right\rangle/N$. As temperature increases, $\left\langle c^{\dagger}c\right\rangle/N$ decreases monotonically until it reaches 0 at the critical temperature $T_c$ which is given by 
\begin{equation}
\frac{4\omega_{0}\omega_{L}}{\tilde{\Omega}^{2}}=\tanh\left(\frac{\omega_{0}}{2\omega_{L}k_{B}T_{c}}\right)\,.
\end{equation}
For the example shown in Fig.~\ref{fig5}(c) with finite $E_r$, the system is in the normal phase with $\left\langle c^{\dagger}c\right\rangle/N=0$ at both the low and the high temperature ends, and is in the superradiant phase in an intermediate temperature window between $T_{c}^{(1)}$ and $T_{c}^{(2)}$.

A remark is in order. In our derivation of the partition function $Z$ in Eq.~(\ref{eq:Z}), we have treated the $N$ atoms as 
distinguishable particles which obey the Boltzmann distribution. In other words, we have ignored the quantum statistics of atoms. This should be a good assumption at high temperature. We may estimate the temperature regime in which this assumption is valid as follows. Let us assume that the atoms are ideal bosons. The critical temperature for the bosons to form Bose-Einstein condensate is given by
\begin{equation}
T_{\rm{BEC}}=3.31\frac{\hbar^{2}\rho^{\frac{2}{3}}}{mk_{B}}\approx3\times10^{-4}\left(\frac{\hbar\omega_{0}}{k_{B}}\right),
\end{equation}
where we have taken typical values such that the atomic number density $\rho=10^{13}\textrm{cm}^{-3}$, $m$
the mass of $\text{{\ensuremath{^{87}}Rb}}$ atom, and energy splitting between two ground state hyperfine states $\omega_{0}=2\pi\times4.81\textrm{MHz}$.
When $T\gg T_{\rm{BEC}}$, quantum statistics is not important, and the bosons can in practice be treated as distinguishable particles. As $T_{\rm{BEC}}$ is very small in our unit system, our results as presented in Fig.~\ref{fig5} should largely remain valid for typical experimental situations. Note that as $T_{\rm{BEC}}$ is roughly the same as Fermi degenerate temperature, this estimate is also valid for a system of Fermi gas. How to properly incorporate quantum statistics of atoms in the calculation for temperatures within the quantum degenerate regime remains a challenge and will be investigated in the future.

\section{conclusion}

In conclusion, we have studied the Raman spin-orbit coupling induced
by one cavity photon field and one classical Raman laser
beam, where all three degrees of freedom including the atomic internal
pseudo-spin, the atomic external COM motion, and the cavity photon field are coupled and treated self-consistently. 
For the single-atom case, we show that the SOC stabilizes the $n=0$ sector which contains no photons. Furthermore, the SOC combined with a trapping potential gives rise to rich spin dynamics. For the many-atom
case in thermodynamic limit, we focused on the physics of the Dicke superradiance
phase transition. In comparison to the TC model where the atomic COM motion is neglected, the SOC 
modifies the phase transition boundary by increasing the critical atom-cavity coupling strength at which the system becomes superradiant. Furthermore, the non-monotonic behavior of the critical coupling strength can lead to the reentrant of the non-superradiant normal phase as the temperature varies.

\textit{Acknowledgment} --- This research is supported by the NSF (Grant No. PHY-1505590) and the Welch Foundation (Grant No.
C-1669).

\appendix

\section{Perturbation Theory for Oscillation Frequency Shift of $P_{\uparrow}\left(t\right)$\label{sec:perturbation-theory-for}}
In this Appendix, we provide a detailed derivation of Eq.~(\ref{eq:f_pertur1}) using a perturbation calculation. It is more convenient to carry out the calculation in the lab frame, in which the Hamiltonian reads
\begin{equation}
h_{t}^{\rm{lab}}=\frac{\hat{k}^{2}}{2m}+\frac{1}{2}m\omega_{t}^{2}x^{2}+\frac{\delta}{2}\hat{\sigma}_{z}+\frac{\Omega_{{\rm cl}}e^{2iq_{r}x}}{2}\hat{\sigma}^{+}+\frac{\Omega_{{\rm cl}}e^{-2iq_{r}x}}{2}\hat{\sigma}^{-}\,,\label{eq:h_t_lab}
\end{equation}
which is the counterpart of Hamiltonian (\ref{eq:h_t_np}).

In the limit of large trapping frequency $\omega_t \gg E_r$, the atoms are tightly confined within a spatial region much smaller than $1/q_r$. Hence we may Taylor expand the two exponentials in Hamiltonian (\ref{eq:h_t_lab}) to second order in $q_r$, and write 
\begin{equation}
h_{t}^{\rm{lab}}=h_{0}+V,\label{eq:H_pertur}
\end{equation}
where
\begin{align}
h_{0} & =\frac{k^{2}}{2m}+\frac{1}{2}m\omega_{t}^{2}x^{2}+\frac{\delta}{2}\hat{\sigma}_{z}+\frac{\Omega_{{\rm cl}}}{2}\hat{\sigma}^{+}+\frac{\Omega_{{\rm cl}}}{2}\hat{\sigma}^{-};\\
V & =\left(iq_{r}x-q_{r}^{2}x^{2}\right)\Omega_{{\rm cl}}\hat{\sigma}^{+}-\left(iq_{r}x+q_{r}^{2}x^{2}\right)\Omega_{{\rm cl}}\hat{\sigma}^{-}.
\end{align}
We shall treat $V$ as a perturbation to $h_0$, and focus on the case with $\delta=0$.

The eigenenergies and eigenstates of the unperturbed Hamiltonian $h_0$
are given by
\begin{align}
E_{q\pm}^{(0)} & =\left(\frac{1}{2}+q\right)\omega_{t}\pm\frac{\Omega_{{\rm cl}}}{2};\label{eq:E_0th}\\
\left|q\pm\right\rangle  & =\frac{1}{\sqrt{2}}\left(\left|\uparrow\right\rangle \pm\left|\downarrow\right\rangle \right)\left|q\right\rangle ,
\end{align}
where $q$ is the harmonic oscillator quantum number. Our initial state has the atom in $|\downarrow \rangle$ and harmonic oscillator ground state $|q=0 \rangle$, which can be written as
\begin{equation}
\left|\psi (0)\right\rangle =\frac{1}{\sqrt{2}}\left(\left|0+\right\rangle -\left|0-\right\rangle \right) \,.
\end{equation}
Neglecting $V$, the ensuing dynamics will lead to Rabi oscillation with frequency $\Omega_{\rm cl}$, i.e., the energy difference between the two eigenstates $|0 \pm \rangle$. This is the result for the JC model.

To find the oscillation frequency when $V$ is included, we shall calculate the energy shift to the states $|0 \pm \rangle$ to second order in $q_r$. The corresponding oscillation frequency will then be 
\begin{equation}
f=\left(E_{0+}^{(0)}+E_{0+}^{(1)}+E_{0+}^{(2)}\right)-\left(E_{0-}^{(0)}+E_{0-}^{(1)}+E_{0-}^{(2)}\right)\label{eq:f}
\end{equation}
with $E_{0\pm}^{(1)}$ and $E_{0\pm}^{(2)}$ being the $1st$ and $2nd$
order energy shift due to the perturbation $V$, respectively. Through
the standard time independent perturbation theory, we obtain
\begin{equation}
E_{0\pm}^{\left(1\right)}=\left\langle 0\pm\right|V\left|0\pm\right\rangle =\mp\frac{E_{r}\Omega_{{\rm cl}}}{\omega_{t}};\label{eq:E_1st}
\end{equation}
and
\begin{align}
E_{0\pm}^{\left(2\right)} & =\frac{\left|\left\langle 1\mp\right|V\left|0\pm\right\rangle \right|^{2}}{E_{0-}^{\left(0\right)}-E_{1\mp}^{\left(0\right)}}+\frac{\left|\left\langle 2\pm\right|V\left|0\pm\right\rangle \right|^{2}}{E_{0-}^{\left(0\right)}-E_{2\pm}^{\left(0\right)}}\label{eq:E_2nd}\\
 & =-\frac{E_{r}\Omega_{{\rm cl}}^{2}}{\omega_{t}\left(\omega_{t}\mp\Omega_{{\rm cl}}\right)}-\left(\frac{E_{r}\Omega_{{\rm cl}}}{\omega_{t}}\right)^{2}\frac{1}{\omega_{t}}.\nonumber 
\end{align}
Substituting Eqs. (\ref{eq:E_0th})(\ref{eq:E_1st})(\ref{eq:E_2nd})
into Eq. (\ref{eq:f}), we obtain the oscillation frequency of $P_{\uparrow}\left(t\right)$
\begin{equation}
f\left(\omega_{t}\right)=\Omega_{{\rm cl}}-\frac{2E_{r}\Omega_{{\rm cl}}}{\omega_{t}}-\frac{2E_{r}\Omega_{{\rm cl}}^{3}}{\omega_{t}\left(\omega_{t}^{2}-\Omega_{{\rm cl}}^{2}\right)},\label{eq:f_pertur}
\end{equation}
as given in Eq.~(\ref{eq:f_pertur1}) in the main text.

\end{document}